\documentclass[acmsmall]{acmart}

\makeatletter

\usepackage{tablefootnote}
\usepackage{graphicx}
\usepackage{subcaption}
\usepackage{booktabs}
\usepackage{adjustbox}

\newcommand{\addedcontent}[1]{\textcolor{black}{#1}}
\newcommand{\deletedcontent}[1]{\textcolor{lightgray}{}}

\newcommand{\Braids}{\textsc{braids.social}}

\newcommand{\pquote}[1]{\textcolor{black}{\textit{``#1''}}}
\newcommand{\dquote}[1]{{``#1''}}

\makeatletter
\newcommand*\bigcdot{\mathpalette\bigcdot@{.5}}
\newcommand*\bigcdot@[2]{\mathbin{\vcenter{\hbox{\scalebox{#2}{$\m@th#1\bullet$}}}}}
\makeatother

\AtBeginDocument{%
  }

\setcopyright{acmlicensed}
\copyrightyear{2018}
\acmYear{2018}
\acmDOI{XXXXXXX.XXXXXXX}
\acmConference[CSCW]{Make sure to enter the correct
  conference title from your rights confirmation email}{Oct,
  2025}{Bergen, Norway}
\begin{document}

%%
%% The "title" command has an optional parameter,
%% allowing the author to define a "short title" to be used in page headers.
\title{Understanding Decentralized Social Feed Curation on Mastodon}

%%
%% The "author" command and its associated commands are used to define
%% the authors and their affiliations.
%% Of note is the shared affiliation of the first two authors, and the
%% "authornote" and "authornotemark" commands
%% used to denote shared contribution to the research.
\author{Yuhan Liu}
\email{yl8744@princeton.edu}
\orcid{0000-0001-6852-6218}
\affiliation{%
  \institution{Princeton University}
  % \streetaddress{35 Olden St}
  % \city{Princeton}
  % \state{New Jersey}
  \country{USA}
  % \postcode{08544}
}

\author{Emmy Song}
\email{etsong@princeton.edu}
\affiliation{%
  \institution{Princeton University}
  % \streetaddress{35 Olden St}
  % \city{Princeton}
  % \state{New Jersey}
  \country{USA}
  % \postcode{08544}
}

\author{Owen Xingjian Zhang}
\email{owenz@princeton.edu}
\orcid{0009-0008-2949-7379}
\affiliation{%
  \institution{Princeton University}
  % \streetaddress{35 Olden St}
  % \city{Princeton}
  % \state{New Jersey}
  \country{USA}
  % \postcode{08544}
}
\author{Jewel Merriman}
\email{jm0278@princeton.edu}
\affiliation{%
  \institution{Princeton University}
  % \streetaddress{35 Olden St}
  % \city{Princeton}
  % \state{New Jersey}
  \country{USA}
  % \postcode{08544}
}

\author{Lei Zhang}
\email{raynez@princeton.edu}
\orcid{0000-0003-3584-6754}
\affiliation{%
  \institution{Princeton University}
  % \streetaddress{35 Olden St}
  % \city{Princeton}
  % \state{New Jersey}
  \country{USA}
  % \postcode{08544}
}

\author{Andrés Monroy-Hernández}
\email{andresmh@princeton.edu}
\orcid{0000-0003-4889-9484}
\affiliation{%
  \institution{Princeton University}
  % \streetaddress{35 Olden St}
  % \city{Princeton}
  % \state{New Jersey}
  \country{USA}
  % \postcode{08544}
}

%% By default, the full list of authors will be used in the page
%% headers. Often, this list is too long, and will overlap
%% other information printed in the page headers. This command allows
%% the author to define a more concise list
%% of authors' names for this purpose.
\renewcommand{\shortauthors}{Liu et al.}

%%
%% The abstract is a short summary of the work to be presented in the
%% article.
\begin{abstract}
As centralized social media platforms face growing concerns, more users are seeking greater control over their social feeds and turning to decentralized alternatives such as Mastodon. The decentralized nature of Mastodon creates unique opportunities for customizing feeds, yet user perceptions and curation strategies on these platforms remain unknown. This paper presents findings from a two-part interview study with 21 Mastodon users, exploring how they perceive, interact with, and manage their current feeds, and how we can better empower users to personalize their feeds on Mastodon. We use the qualitative findings of the first part of the study to guide the creation of \Braids, a web-based prototype for feed curation. Results from the second part of our study, using \Braids, highlighted opportunities and challenges for future research, particularly in using seamful design to enhance people's acceptance of algorithmic curation and nuanced trade-offs between machine learning-based and rule-based curation algorithms. To optimize user experience, we also discuss the tension between creating new apps and building add-ons in the decentralized social media realm. 
\end{abstract}

%%
%% The code below is generated by the tool at http://dl.acm.org/ccs.cfm.
%% Please copy and paste the code instead of the example below.
%%


%%
%% Keywords. The author(s) should pick words that accurately describe
%% the work being presented. Separate the keywords with commas.
%% A "teaser" image appears between the author and affiliation
%% information and the body of the document, and typically spans the
%% page.
\begin{comment}
\begin{teaserfigure}
  \includegraphics[width=\textwidth]{Figures/teaser.pdf}
  \caption{We conducted a two-part interview study on decentralized social media, proposed the design space for feed curation in Mastodon, and built \Braids, a web client that enables users to customize their feeds in Moastodon.}
  \Description{In this paper, we conducted a two-part interview study, proposed the design space for feed curation in Mastodon, and built \Braids, a web client that enables users to customize their feeds}
  \label{fig:teaser}
\end{teaserfigure}
\end{comment}

\received{20 February 2007}
\received[revised]{12 March 2009}
\received[accepted]{5 June 2009}

%%
%% This command processes the author and affiliation and title
%% information and builds the first part of the formatted document.
\maketitle

\section{Introduction}
%background and research questions
\addedcontent{Social feeds play a crucial role in shaping online discourse, influencing what information users see, engage with, and believe~\cite{diakopoulos2014algorithmic, bakshy2015exposure, lerman2010information}. Mainstream centralized social media platforms have implemented sophisticated algorithmic curation in their social feeds to customize content to individual interests and behaviors \cite{shirky2011political, rodriguez2014quantifying}. In the past decade, HCI and CSCW researchers have investigated social feed curation in mainstream centralized social media platforms from multiple perspectives such as content moderation~\cite{vaccaro2021contestability, feuston2020conformity}, algorithm transparency~\cite{eslami2015always, eslami2016first}, and interface design~\cite{feng2024mapping, eslami2015feedvis, he2023cura}. Recently,} Mastodon \addedcontent{emerged} as a decentralized alternative to mainstream centralized social media platforms, \addedcontent{featuring chronological feeds without AI-driven curation~\cite{joinmastodon}. It }has attracted considerable attention and a growing user base following Twitter's ownership and policy changes in 2022 ~\cite{twitter_mastodon_bloomberg, twitter_mastodon_mit_review, cava2023drivers}. On the surface, Mastodon shares many similarities with Twitter, such as short text posts and re-posts ~\cite{jeong2023exploring, LACAVA2022100220, raman2019challenges}.
However, it operates \addedcontent{fundamentally }\deletedcontent{very }differently: \addedcontent{unlike Twitter’s centrally controlled infrastructure,} Mastodon is a decentralized network of servers independently administered by individuals or organizations that interoperate with one another using the ActivityPub protocol~\cite{anaobi2023will}, but lack ``universal'' moderation policies ~\cite{lee2023uses}.

Social feeds on Mastodon are also different from centralized social media platforms because they are partly sourced from the database of the user’s ``home server''\footnote{The home server is the server where a user’s account is hosted}, which consists of posts from users on the same server and posts from a mix of server it connects with, rather than one centralized database. Moreover, the organization that develops the Mastodon open-source software and operates the largest server is a non-profit that claims to prioritize user agency and preference over corporate interests or ad revenue ~\cite{masnick2019protocol, joinmastodon}. \addedcontent{They open-sourced the source code of web and mobile client, together with the API}. This opens up opportunities for developing user-centered feed curation tools.
%3) The non-profit structure of Mastodon allows for feed curation that prioritizes user preferences over corporate interests or ad revenue ~\cite{masnick2019protocol}; 4) the open-source nature of Mastodon enables the development and integration of third-party tools for feed customization and moderation. 
%5) Most of the feed curation tools in centralized social media platforms are data-driven~\cite{bernstein2010eddi, bernstein2010torrent, duh2012creating, he2023cura, gilbert2012predicting}, but the relatively smaller size of the database of each server opened up opportunities for rule-based curation algorithms.

\addedcontent{Therefore, based on the differences in the operational model and technical architecture between decentralized social media like Mastodon and mainstream centralized social media platforms, we highlight key opportunities for understanding and designing feed curation tools for Mastodon users. First, the independence of each server results in varied content across the decentralized network, which leads to information sharing and consumption expectations that differ from those of mainstream centralized social media users. For instance, recent studies have revealed that the main motivations for using Mastodon are social escapism and discussing identity-based topics within a safe community ~\cite{lee2023uses, zhang2024trouble}. These varying expectations may influence user preferences in feed curation, making it essential to first investigate users' perceptions and interaction patterns before developing design solutions. Secondly, Mastodon's decentralized nature introduces a unique user experience in curating social feeds, especially in configuring feeds from multiple data sources, which necessitates exploration in different interface designs for prioritizing transparency and user control and adaptability to the decentralized structure. }\deletedcontent{While previous research on feed curation tools has primarily focused on centralized platforms~\cite{bernstein2010eddi, bernstein2010torrent, duh2012creating, he2023cura, gilbert2012predicting}, this paper aims to understand Mastodon users' perceptions and needs for feed curation. Specifically, }\addedcontent{In this project}, we seek to answer the following research questions:
\begin{itemize}
        \item RQ1: How do people perceive, \deletedcontent{and }interact with \addedcontent{and manage} decentralized feeds in Mastodon?
        % \item \deletedcontent{RQ2: How do people manage decentralized feeds in Mastodon?}
        \item \addedcontent{RQ2: What interface design facilitates explicit preference expression for feed curation in Mastodon?}
    \end{itemize}
% Feed curation in traditional centralized social media platforms has been widely studied in the past decades. 

To answer these research questions, we conducted a two-part interview study with 21 participants. In part one of the study, we asked participants how they perceive, interact with, and manage their current feeds. Qualitative findings showed a common preference for chronologically ordered feeds, but participants are open to certain algorithmic curation. We also noticed the under-utilization of certain feeds due to the information overload. Based on the findings, we built \Braids{}, a web-based feed curation tool that collapses all feeds into one semi-chronological and uses a slider-based interface to collect user preferences. We then conducted the second part of the study, during which participants were asked to configure their unified feed with \Braids{}. %We explored their perceptions and opinions about affordance design and further understood their preference for feed curation in Mastodon. 
Lastly, we highlighted the design implications and insights we gained through the study for future research in the decentralized social media realm. 

In summary, we made the following contributions:
\begin{itemize}
    \item We investigated users' perception and interaction with decentralized feeds in Mastodon.
    \item We built \Braids{}, a web-based tool for feed curation on Mastodon that investigates users' preferences when interacting with feed curation tools.
\end{itemize}

\section{Related Work}
\subsection{Decentralized Social Media and Mastodon}

Decentralized social media has emerged as an alternative to growing concerns associated with centralized platforms, gaining even more popularity after Elon Musk's Twitter acquisition (later renamed to X)~\cite{twitter_mastodon_bloomberg, twitter_mastodon_mit_review, nicholson2023exploration}. Unlike centralized social media platforms, which are typically managed by a corporation, decentralized social media represents a user-controlled ecosystem with distributed information management~\cite{datta2010decentralized}. They often employ standardized protocols for data exchange. For example, Mastodon adopts the ActivityPub Protocol~\cite{PrincetonDSM2024, mansoux2020seven} and Bluesky implements the ATProtocol~\cite{bluesky}. People often use the term \dquote{Fediverse} to describe the ensemble of social networks that can communicate through decentralized protocols~\cite{wiki:fediverse}. \addedcontent{By the end of Feb 2025, \dquote{Fediverse} has over 15 million users and 22 thousand servers~\cite{Fedidb}}. Beyond differences in technical infrastructure, decentralized social media often offer unique user experiences in data management, privacy control and content moderation,distinguishing them further from their centralized counterparts~\cite{hwang2023whose}. 

Mastodon is one of the most well-known decentralized social media platforms. Like Twitter, Mastodon functions as a microblogging platform where people can publish original posts and boost other people's posts~\cite{la2021understanding, zignani2018follow}. Thanks to the ActivityPub Protocol, when users post on Mastodon, the content is not confined to their \dquote{home server} but is shared across any federated servers~\cite{la2021understanding, activitypub}. \addedcontent{A home server is the server where a user's Mastodon account is hosted. During onboarding, newcomers must select a server to create their account. To simplify this decision and reduce the effort required to find a suitable server, Mastodon provides a curated list of recommended servers on its official website~\cite{joinmastodon}. Alternatively, users can manually search for servers.} Mastodon has four feeds: \dquote{home}, \dquote{local}, \dquote{federated}, and \dquote{explore}. The home feed is personalized, featuring content from the accounts or hashtags the user follows. The three feeds other than the explore feed in Mastodon are organized in chronological order, whereas the explore feed is sorted according to a \dquote{trending score}. The composition of content in each of these feeds is as follows:
\begin{itemize}
    \item \textbf{Home:} Activities from accounts the user follows, including posts, boosts, and replies, together with posts that contain certain hashtags the user follows. 
    \item \textbf{Local:} Public posts from all users hosted by the user's home server.
    \item \textbf{Federated:} All public posts fetched by the server, including posts from accounts followed by local users\footnote{\dquote{Local} user refers to all users hosted by the given server} and the posts that local users have boosted, favorited, replied to and searched for. 
    \item \textbf{Explore:} Trending posts in the federated feed ranked based on the \dquote{trending score}. The \dquote{trending score} is calculated based on time and interaction (boosts and favorites) the post received.
\end{itemize}

\addedcontent{In a word, what a user sees in their home feed is based on the user's subscription in the federated network, while the other three feeds are the expansion of the user's home feed but also depend on the moderation and federation policy of the home server. }In this paper, we contribute to the feed curation mechanism of decentralized platforms by providing empirical insights on users' perception and interaction with Mastodon feeds, and by developing a new curation tool, i.e., \Braids{}, that enables users to customize their feeds.

\subsection{Feed Curation in Centralized social Media Platforms}
As online social media has made sharing information easier, they have also contributed to information overload~\cite{bernstein2010eddi, bernstein2010torrent, shirky2011political, rodriguez2014quantifying}. In the early days, platforms like Twitter and Facebook displayed content chronologically, allowing users to view posts in real-time order. However, because of the massive growth in user bases and content volume, millions of active users generate vast amounts of content daily, overwhelming users with a continuous, unfiltered stream of posts~\cite{bernstein2010eddi, bernstein2010torrent}. To mitigate this, centralized social media platforms developed sophisticated algorithms to curate social feeds for users \cite{shirky2011political, rodriguez2014quantifying}. For example, Twitter, Instagram, and TikTok use AI-driven methods for selecting, ranking, and recommending posts in users' feeds based on interaction prediction~\cite{UnderstandingSocialMediaAlgorithms, wong2018facebook, milli2021optimizing, covington2016deep, meyerson2012youtube}. The content recommendation algorithms have become a key feature of these platforms and are credited with significantly increasing the platforms' user base, reportedly attracting millions of new users ~\cite{bandy2021more, TwitterEngineering2023, dujeancourt2023effects}. 

Here we summarized the main difference in feed curation between centralized social media platforms and Mastodon:
\begin{enumerate}
    \item \textbf{Data source:} The ActivityPub protocol and the federated architecture of Mastodon allow content to be distributed throughout the network when people subscribe to it~\cite{la2021understanding, activitypub}. The database of each server depends on the subscription of each user hosted by the server and the server administrator's federation choice\footnote{Server administrator can decide \dquote{de-federated} with \addedcontent{specific} \deletedcontent{certain} server to stop data exchange}. Therefore, the data used to generate each user's feed is different. 
    \item \textbf{User expectation:} Unlike centralized social media platforms \addedcontent{such as Twitter and Facebook}, Mastodon servers operate independently without a \dquote{universal} moderation policy. Server administrators and moderators decide what is allowed on the server, which varies between different servers, making it possible for Mastodon users to find a server aligned with the content and values they seek. \addedcontent{Also, prior research found that Mastodon users are interested in joining smaller communities in seek of a safe environment for discussion or social escapism~\cite{lee2023uses, zhang2024trouble}. The differences in user expectations may influence their curation needs. }
    %Prior research shows that several main motivations for using Mastodon including adult information sharing and consuming, social escapism (expressing feelings without causing unnecessary social attention)~\cite{lee2023uses}, which is different from what centralized social media platform users~\cite{ruggiero2000uses, papacharissi2000predictors, dholakia2004social}. 
    \item \textbf{Goal of feed curation:} Mainstream centralized social media platforms, funded by venture capitalists, seek ultimate control over the platform to be much better positioned to monetize the platforms via targeted advertising~\cite{masnick2019protocol}. Thus in the development of feed curation algorithms, they prioritize features that can maximize user retention such as the time spent on the platform or the tendency to interact with the post~\cite{UnderstandingSocialMediaAlgorithms, wong2018facebook, milli2021optimizing, covington2016deep, meyerson2012youtube}. Those features do not consider users' preferences. However, the non-profit nature of Mastodon grants user greater agency in curating what they want to see in their feeds. 
    \item \textbf{Allowance of third-party development:} Recently, centralized social media platforms stopped opening their API to the public, which makes the development of third-party curation tools difficult~\cite{masnick2019protocol, mashable2021twitterapi, martech2020facebookapi, graber2021ecosystem, surve2024decentralization}. However, the openness of Mastodon encourages the development of third-party clients, giving options for end-users to have different user experiences based on personal preference~\cite{feditips2023apps}.
    \item \textbf{Data-driven vs. rule-based curation tools:} Prior work has focused on designing data-driven solutions on centralized social media platforms to mitigate information overload and enhance people's network building~\cite{bernstein2010eddi, bernstein2010torrent, duh2012creating, he2023cura, gilbert2012predicting}. Rule-based algorithms, even though they provide more transparency, are neglected due to their limitations in dealing with large amounts of data. However, the relatively smaller size of the database of each server in Mastodon opened up new opportunities for developers. 
\end{enumerate}

People's feelings towards algorithmic curation in centralized social media platforms are mixed. On one hand, people like the personalized experience and efficient information filtering that these algorithms offer\cite{berkovsky2015personalized}. Despite the popularity of these algorithms, the resistance to algorithmic curation has never stopped ~\cite{devito2017algorithms, bandy2021more}. Researchers have investigated the reasons behind the resistance and attribute it to the violation of expectations, misunderstanding of the algorithms, lack of transparency and agency, etc ~\cite{devito2017algorithms, rader2015understanding, eslami2015always, eslami2016first}. Meanwhile, several strategies have been proposed to address these issues; for instance, seamful design not only increases user awareness of algorithms but also improves the quality of human-algorithm interactions and user agency in complex systems ~\cite{eslami2015always, eslami2016first}. 
In this paper, we extend the line of work by exploring the understudied areas about enhancing transparency and user agency of curation tools on decentralized social media.

%Though people's perceptions and preferences in traditional centralized social media platforms are well-studied, the differences between Mastodon feeds and feeds in traditional centralized platforms are different. There is a need to investigate whether the previous findings still apply to the emerging context. We aim to address the research gap in the exploratory study.  
\subsection{Existing Feed Curation Tools in Mastodon}
In this section, we examined five popular Mastodon clients about their unique features (we compare them with the native Mastodon app to define \dquote{unique}) that enable users to curate and customize their feeds when browsing.
\begin{itemize}
    \item Tusky is a Mastodon client for Android users that allows users \dquote{mute} a hashtag if they don't want to see about a certain topic. Tusky also offers a filter that can hide boosts~\cite{TuskyApp}. 
    \item Elk is a web-based Mastodon client that reorders posts connected to each other to display them in a connected thread in the feeds (in chronological feeds sometimes fragment discussions by presenting posts solely in time-based order)~\cite{ElkApp}. 
    \item Ivory is a Mastodon client for iOS and Mac users. It will suppress duplicate boosted posts for users~\cite{IvoryApp}. 
    \item Ice Cube is a Mastodon client for iOS and Mac users. Besides one local feed, users can fetch a specific remote server's local feed without registering an account. Ice Cube also marked where the user stopped and can resume the next time when the user logs in~\cite{IceCubesApp}.
    \item Fedilab is a client for Mastodon and other platforms that integrate the ActivityPub protocol such as Pleroma and Friendica. It supports multiple account management. Users can log in to multiple accounts and switch between different feeds for those accounts simultaneously~\cite{FedilabApp}.
\end{itemize}

Feed curation in these existing Mastodon apps is either on an individual post level, like hiding or suppressing specific posts, or on the feed level without manipulating content in the post, like fetching the entire feed from remote servers. The exception is Elk, which re-orders posts to form a connected thread based on replying time. However, Elk manages feeds in an automatic format without users' input. In this paper, we aim to explore how feed curation can occur across different feeds and better integrate user preferences. 

\section{Method}
\subsection{Study Methodology}
To investigate people's perceptions and opinions regarding Mastodon feeds, \deletedcontent{\textbf{(RQ1) }}their preferences in managing their Mastodon feeds \textbf{(RQ1)} and what are their needs in curation tools \textbf{(RQ2)}, we conducted a two-part interview study with 21 Mastodon users. 
% In part one, we asked participants about their experiences using Mastodon, especially about different feeds, including the frequency they check this feed, their perceptions of the content, and their interaction patterns. Based on our findings, we mapped a design space and built \Braids, a web-based prototype that enabled users to curate their own feed and invited participants to interact with the prototype.

\textbf{Part 1: Understanding People's Perception and Interaction with Mastodon Feeds.} We invited eleven participants (P1-P11) to have 60-minute interview sessions with us. We started by asking participants about their most recent experiences using Mastodon, and their rationale for picking their home server. Then, we invited them to share their screen as we asked them separate questions for each Mastodon feed, including the frequency with which they check this feed, their perceptions of the content, and their interaction patterns. We also asked participants to compare their use of various Mastodon feeds.

\textbf{Part 2: Exploring How to Enable Users to \deletedcontent{Customize their Mastodon Feeds.} \addedcontent{Explicitly Indicate Their Preference in Mastodon Feed Curation.}} We designed and built \Braids{} based on our findings of the first part of the study (details of design consideration and system implementation are presented in the next section following the findings about RQ1\deletedcontent{and RQ2}) and had 30-minute interview sessions with another ten participants (P12-P21). During the interview, we presented \deletedcontent{the design of }\Braids{} and asked participants to complete the following tasks of feed customization \deletedcontent{to explore their opinions about the affordance design in the system}\addedcontent{so they could  explore and learn how to use the system}:

\begin{itemize}
    \item Task 1: See almost entirely trending posts.
    \item Task 2: See fewer posts from users and hashtags they follow relative to the local server. 
    \item Task 3: See twice as many posts from the user @Mastodon on the mastodon.social server as trending posts.
\end{itemize}

\deletedcontent{As participants completed the tasks, we asked them to think aloud as we observed their behavior.} Afterward, we asked them to \addedcontent{think aloud while configuring} \deletedcontent{configure}the feed based on their preferences, followed by questions regarding how they felt about specific features of \Braids{} and how they would like to change or improve it.

\subsection{Participants}
We recruited participants by posting on the home server of the leading author. Subsequently, the recruitment post and its boosts were distributed across the Mastodon network. The recruiting post consisted of a brief introduction to the study and a link for signing up. We conducted semi-structured interviews with Mastodon users via Zoom. Participants' basic information is provided in the Appendix \ref{appendix:participants}. Our university's Institutional Review Board approved this work. Additionally, all participants provided consent to participate in the interviews, to the recording of their voices and screens, and to having their conversations analyzed. \addedcontent{Participants in part one and part two of the study were completely distinct.}

\subsection{Data Analysis}
For part one, we performed thematic analysis to gain deep insights into participants' experiences. Two researchers collaboratively created a 48-code codebook (see Appendix~\ref{tab:part_one_codebook}) from the initial two transcripts, aligning with our research questions and using thematic analysis methods ~\cite{braun2006using}. We validated each code category utilizing a subset of transcripts, categorizing based on similarities in participant responses. Coders met regularly to ensure high inter-rater reliability (Cohen's kappa $\geq$ 0.8) in the coding process. After coding was finished, the leading author iterated on extracting themes from codes. For part two, one export coder performed thematic analysis to code the transcripts and extract themes~\cite{braun2006using, mcdonald2019reliability}. The codebook consists of 40 codes and is provided in Appendix~\ref{tab:part_two_codebook}.
% \input{Sections/3_formative_study}
% \input{Sections/3.5_Design_Space}
% \input{Sections/4_braids}
% \section{Results}
\section{\addedcontent{Findings: Understanding People’s Perception and Interaction with Mastodon Feeds. \deletedcontent{RQ1: What do people think of Mastodon feeds and how do they interact with them?}}}
\subsection{Participants prefer chronological feeds but also identified  scenarios where algorithmic feeds might help}
Mastodon distinguishes itself from other social media platforms because its feeds (except for the explore feed) are ordered chronologically, often referred to as ``not using an algorithm'' by its users. When participants talked about algorithmic feeds, they expressly referred to the machine learning algorithms that rank posts based on maximization of user engagement or time spent on the platform. 

Most participants (8 out of 11) liked the chronological order of Mastodon's feeds for the transparency it grants. For example, P2 likes the chronological order as it is not manipulated and clearly reveals what she missed while away: \pquote {I love that it's chronological. I just want to see everyone I am following and nothing more. When I was away for 12 hours, I could catch up on that 12 hours of content without having to fiddle with it and be like, `Okay, where is all this stuff now?' I just want to see exactly what I missed.} 

Participants drew comparisons between Mastodon's chronological feed order and the algorithmic recommendation on mainstream social media like Twitter or Facebook, expressing a clear dislike of that kind of algorithmic feed. We then learned that their reluctance was not to all algorithms. Rather, they were critical of the aforementioned machine learning-based algorithms, which did not meet their needs and made them feel powerless. For example, P11 indicated her frustration with the algorithmic feeds on Twitter but noted that she found the algorithm on Bluesky, another decentralized alternative, more satisfactory: \pquote{I love it (chronological order). It's the best. I mean, it's one of the very best things about Mastodon. I hate the whole algorithm thing. [...]  Bluesky doesn't irritate me as much as Twitter or others because at least you can, and they also give you the freedom to create an algorithm.}

% Another apprehension regarding algorithm-curated feeds is the fear that they could be manipulated, compromising their fairness. Mastodon's \textit{explore} feed selects content using a scoring system to rank trending posts, hashtags, and news links. The details of the ranking algorithm are accessible in Mastodon's open-source repository, but it is written in Ruby, which means that only a few users with technical skills can fully understand it. P9 told us her concerns about the potential for abuse by individuals equipped with those skills: \pquote{So my concern about explore (feed) is that it might give a skewed for someone who doesn't understand how the explore (feed) is constructed. It might give a skewed idea of who the heavy hitters are on Mastodon. [P9, as she pointed to an account frequently shown in her explore feed and explained that the owner of the account is a software expert]}

Some participants stay neutral, weighing the pros and cons of both chronological and algorithmic feed curation methods, like P1 and P8: \pquote{I miss that algorithmic mediation where it would still push the most relevant stuff to the top and then keep this like irrelevant kind of stuff at the bottom. [...] But then the chronological (order), works in a way that I don't really miss out on stuff. [...] At least on Mastodon, I know if you put something, it's going to show up for me because of the chronological timeline. Like, I'm not gonna lose it in the algorithmic ordering. So I see both pros and cons,} (P1).\pquote{I like that there's no algorithm that trying to get me to see certain posts and make others harder to find. The algorithm does have the advantage if I've been away for a long time, of helping me kind of sift through,} (P8).

P6 elaborated on his vision for what an ideal algorithmic curation would work from his perspective:\pquote{I understand that algorithmic timelines \footnote{In Mastodon, users often use the term \dquote{timeline} to refer to feeds. We kept the word \dquote{timeline} in quotes and used the feeds in other parts of the paper to be consistent with previous work in this field} are, you know, people don't like that idea. [...] But if I can design my own, or just teach my own algorithm, it'd be great. It doesn't have to have anyone else's input. ... Every time I see his post, I can say, I want to see more of this, I want to see more of that to make sure that the algorithm would say, `Okay, every time this person posts I'm going to make sure it's pinned at the top of his timeline and he sees it as the first thing he sees as soon as he comes back to Mastodon.'} 

In summary, we found that while most participants clearly preferred chronologically ordered feeds, they were not completely against algorithms. In fact, we learned that people find chronological ordering most valuable when they check their feeds frequently, as it provides a timely overview of activities during their absence. However, they identified some scenarios in which algorithm intervention can be useful. For example, when they haven't checked their feeds for longer, they express a desire for algorithmic recommendations to highlight missed content tailored to their interests. They disliked the machine learning algorithmic recommendations due to the lack of agency and transparency.

\subsection{Participants selectively interact with Mastodon feeds by choosing the home server based on community and moderation policy}
One of the most important things for Mastodon users is to choose a home server to host their account. This decision determines what a user sees in their local and federated feeds. In our study, we found a similar pattern that previous studies have argued: people on social media tend to gather because of shared identity, culture, or interests ~\cite{wilkinson2008strong}. For example, P9 joined her home server because that's a community for people of color, while P1 was drawn to join an academic-focused server: \pquote{When I hit Mastodon, I'm just like, `all of the black people come in.' We are like, `Are there black people here?'} (P9). Another motivation for participants to join their home server is to get more information about their local communities. As P10 stated: \pquote{It seemed like a good sort of local/regional instance. [...] There was a group of [the region where P10 lives] people who seemed to join up around that same time as me. And that's where some other people I knew landed. So seemed like a good fit.}

However, some participants are concerned that prolonged engagement within a single community holding similar viewpoints can lead to the formation of echo chambers, as P4 stated \pquote{I specifically chose the larger one to see a cross-section of people because Mastodon is kind of echo-chambered in a way that the same kinds of people I'm finding are like me in many ways. I didn't want to be fully echo-chambered into just a science silo or something like that. So it was a deliberate choice to go sort of broader than more narrow.} As a matter of fact, the communication mechanism of Mastodon enables the user to connect with their preferred community no matter where their account is hosted. Therefore, people do not feel the necessity of joining a specific server. Joining a general-purpose server can expose them to other communities while keeping them in touch with their interested ones. For instance, P6 initially joined a large general-purpose server during the onboarding process. Even though tech-focused communities align more closely with his interests, he expressed no intention to join: \pquote{I know there are other tech instances programmers use. [P6 listed several tech instances] But I don't know what that would necessarily get me because I just need to follow these people. And I also don't want to be completely isolated from other content.}

Moreover, as stated before, Mastodon's servers operate with their own rules and moderation policies. The choice of home server not only determines the content appearing in their federated and local feeds but also reflects the code of conduct they agree to follow.
Among the eleven participants, six highlighted their fondness for what they perceived as a friendly atmosphere in Mastodon, which is largely attributed to users' ability to select a server with moderation policies that resonate with their values, as P3 stated: \pquote{Picking your home server is probably the single biggest stumbling block people have to use Mastodon because there isn't one big site to use. You have to pick one. And that introduces friction. What I pick as my home should be one where I agree with the moderation rules. The people seem generally chill looking at the ABOUT page before you join the server.}

However, finding such a server may take time. Three of our participants (P2, P3, and P7) switched their home servers to find moderation more in tune with their values. \pquote{I switched to this server because I felt that [P2's home server] was moderating, really, really well.} (P2, who used to be on a general-purpose server)

In sum, the home server on Mastodon significantly influences the content they encounter in their local and federated feeds. The primary factor guiding users in choosing their home server is the community they wish to engage with. These choices often revolve around identity,  interests, and geographical location. However, such a choice is not obligatory due to the inter-server connectivity of Mastodon and the Fediverse more broadly. This allows users to communicate and interact with people across different servers, thereby offering flexibility and diversity in their user experiences on Mastodon. Another factor that drives people's choice of home server is the moderation policy. Participants tend to choose the home server whose moderation policy aligns with their values.

\subsection{Participants' interaction frequency of Mastodon feeds depends on their home server size}
Generally, participants reported visiting their home feed the most, followed closely by local and then explore and federated feeds. This pattern is shaped by both the nature of the feeds and the size of the servers. 

Participants indicated that the content in the federated feed is often overly diverse and seldom aligns with their interests. Moreover, the fast pace at which the content pops up contributes to a sense of overwhelm, as P2 stated: \pquote{The federated timeline is so big and wide, which is great, but I have enough things to see. I was like: `Please don't get me more stuff.} [As P2 was showing their federated feed on her screen refreshing with new posts]. For some of the larger servers, even the local feed can appear chaotic, making it challenging for P6 to find content that genuinely interests him: \pquote{It's really all about being able to find the stuff that I'm interested in, and the local and federated timelines for me are just too overwhelming and too general, there's too much content.} To avoid this, P8 joined a relatively small server: \pquote{(My home server) is a relatively small instance. It's a good place to just get to know a couple of people without being overwhelmed with the flow on local timeline.}

Nonetheless, the federated feed is not always useless. About half of our participants (P1, P2, P3, P7, P9) reported that when they are in search of intriguing posts or discovering new follows, they deliberately visit the federated feed. This is particularly true for those new to Mastodon and who have not yet fully built their home feed. The methods these participants use to curate their home feed will be discussed in the following section.\\
\deletedcontent{\textbf{5 \quad \quad RQ2: How Do People Manage the Feeds They See on Mastodon?}}

\subsection{Participants manage their home feed by applying filters for hashtags or accounts}
The chronological order in Mastodon feeds ensures complete transparency in feed organization, granting users full control over the content of their home feeds. In other words, only users' follows and subscriptions can appear in their home feed. This section delves into the strategies and methods users employ to curate their home feeds.

\begin{figure}[t!]
  \centering
  \includegraphics[width=0.7\linewidth]{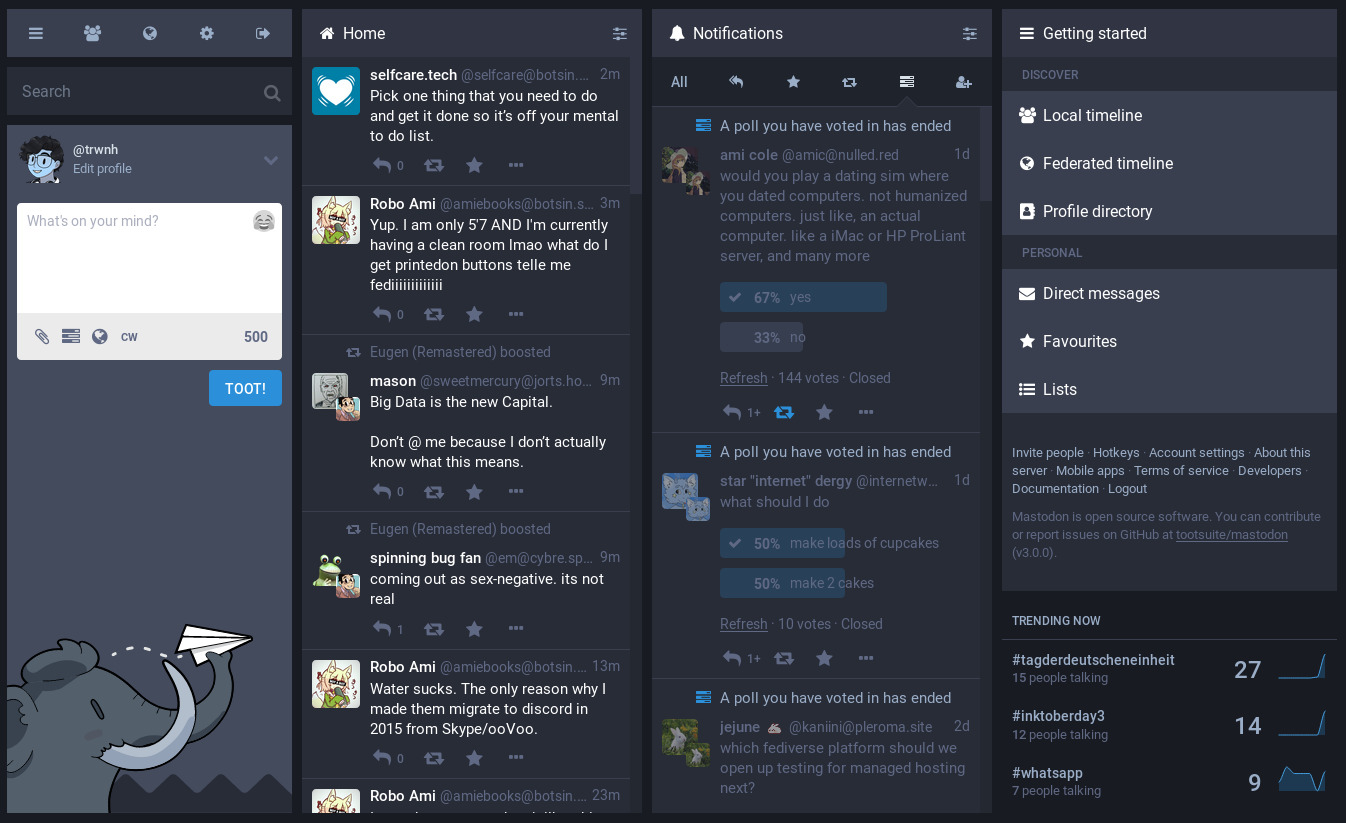}
  \caption{Advanced Web Interface in the Official Mastodon Web App. }
  \label{advanced-interface}
  \vspace{-1em}
\end{figure}

P3, P4, and P11 mentioned using the advanced user interfaces in the native Mastodon web app, which looks similar to TweetDeck~\cite{sump2012making}. This interface lets users view multiple feeds in separate columns on their screens, as shown in Figure \ref{advanced-interface}. By default, columns for the home feed and notifications are displayed, but people can add columns to the layout.

%(\url{https://docs.joinmastodon.org/assets/advanced-web-ui.jpg}).

In addition to following Mastodon accounts, participants indicated that they also follow a variety of hashtags that align with their interests, such as \#cats and \#catsofmastodon. When a user follows a hashtag, all public posts featuring that tag will appear in the user's home feed, regardless of whether or not they follow the account.

We learned that participants apply filters on their home feeds, extracting specific contents to create new columns in the advanced user interface. For example, participants noted that they aggregated selected hashtags and then pinned them to the advanced web interface as a new column. Similarly, participants mentioned using Mastodon's lists feature to create a subset of their followed users. This feature allows users to add specific accounts to lists, creating a new feed exclusively populated by activities from users within that list.

P6 shared with us that when he and the accounts he follows were in different time zones, posts from them could easily be overlooked if he didn't sift through all the posts accumulated during his offline periods. To address this, he set up notifications for specific accounts he follows. This ensures that he receives an alert whenever these specific accounts post, enabling him to stay updated even if they are not online concurrently. 

P9 uses a third-party service and could exclude all boosts from her personal page. P10 told us he uses a third-party app that supports local-only posts. The local-only feature allows people to post on the local server without spreading to the Fediverse. P10 sometimes uses the software to view local-only posts because \pquote{it is local-oriented content or possibly even (related to) someone I know.} The two methods mentioned above, as well as adding hashtags to form a new feed and grouping followed accounts into \dquote{lists}, all fundamentally revolve around applying filters on existing feeds. These methods can be synthesized into rule-based algorithms, aligning with our earlier observation that people are open to algorithmic feed curation. 

\subsection{Participants use multiple apps/tools at the same time because there is no one-fits-all solution}

The adoption of the decentralized protocol ensures multiple clients on the market that are compatible with Mastodon besides the official app Mastodon developed ~\cite{masnick2019protocol}. It is noteworthy that users presented a preference for third-party clients. Seven out of eleven participants said they like using third-party clients (e.g., Elk, Fedilab). One of the advantages of ActivityPub-based social media is that they don't need to install different apps for each platform as they all implement the same protocol and can communicate with each other. However, we learned that participants use multiple apps and tools simultaneously instead of using one. Each client provides unique features that satisfy participants' needs and cannot be replaced. 

P2, a participant with visual impairments, prefers using `Tweezecake' due to its compatibility with screen readers, as it does not necessitate keeping multiple windows open. However, `Tweezecake' falls short in effectively managing hashtag following, prompting her to use `Mona,' a Mastodon client for customized home feed organization. Despite this, `Mona' lacks the feature to synchronize settings across different devices, leading her to adopt `Tuskey' for this specific function. While P2 is comfortable using separate clients for her computer and mobile phone, she finds the necessity of using both `Mona' and `Tusker' on her cellphone cumbersome: \pquote{I'm happy to have two, one for my computer, one for my phone. I just hate that I have to use 2 different apps on my phone to really get what I need. It feels very clunky sometimes.}

P2's situation is not unique; participants P4, P9, and P10 also rely on multiple applications for Mastodon. However, given the limited number of interviews we did, it remains inconclusive whether using multiple apps is more advantageous than a single app equipped with various plugins.

\section{Design Considerations and System Implementation of \Braids{}}

\begin{figure}[hbt!]
    \centering
    \includegraphics[width=\linewidth]{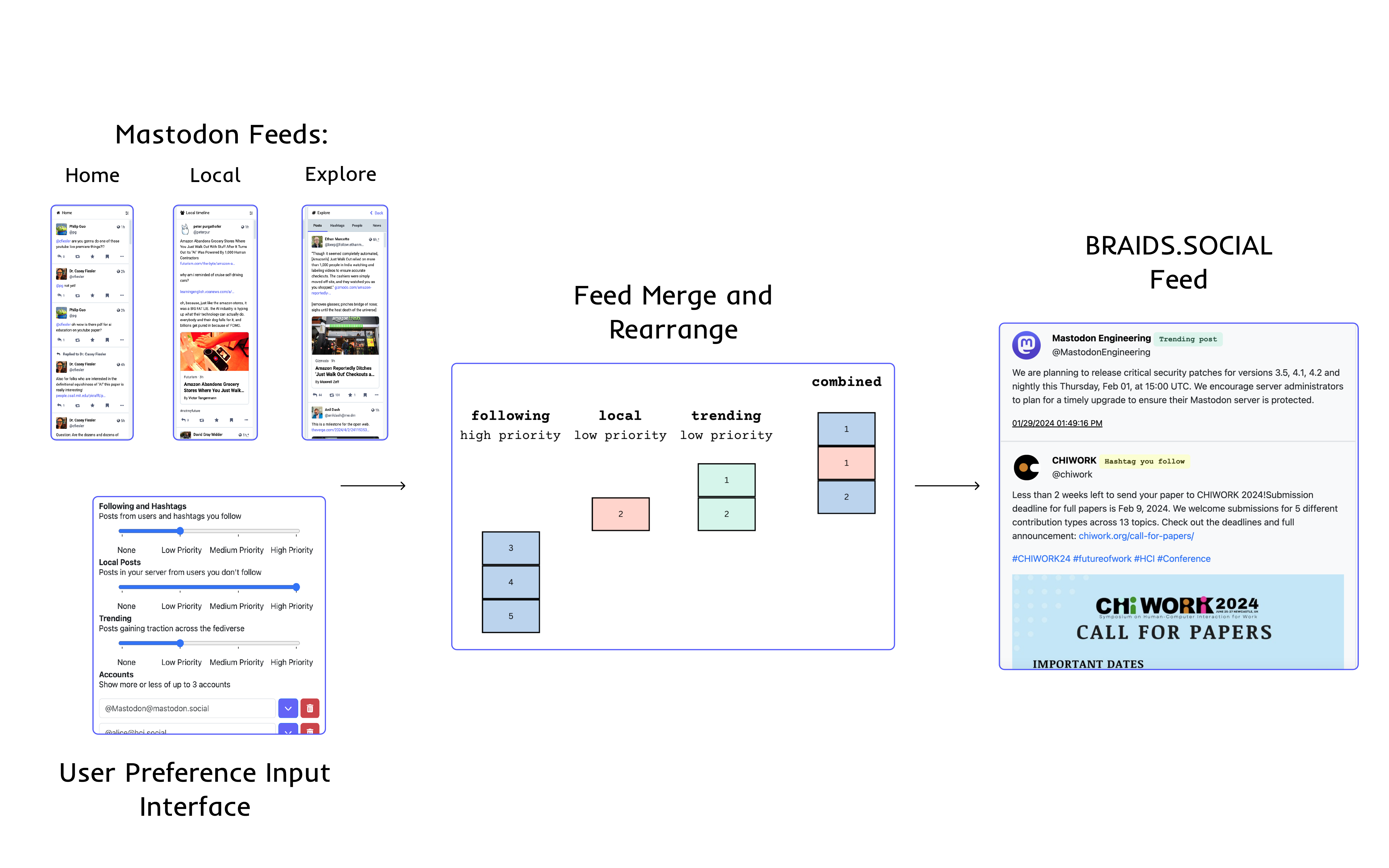}
    \caption{An overview of how \Braids{} generate feed}
    \label{fig:braids}
\end{figure}

\addedcontent{In part one study, we learned that participants are open to algorithmic curation as long as it highlights transparency. Meanwhile, the tools participants currently use for browsing Mastodon feeds do not have algorithmic curation. In part two of the study, we aim to explore the interface design that can help Mastodon users configure rule-based algorithmic curation. }\deletedcontent{To answer this question, }We built \Braids{} based on our aforementioned findings and used it to investigate participants' needs in the tool that helps them customize their feeds. An overview of how \Braids{} generates feed is shown in Figure \ref{fig:braids}. In this section, we \deletedcontent{first} present the design considerations and implementations of \Braids{}, and we report our qualitative findings \addedcontent{in the following section}. 

\subsection{\addedcontent{Design Implications Informed by Part One Study}}
\begin{figure}[t!]
    \centering
    \begin{subfigure}[t]{0.32\textwidth}
        \centering
        \includegraphics[width=\linewidth]{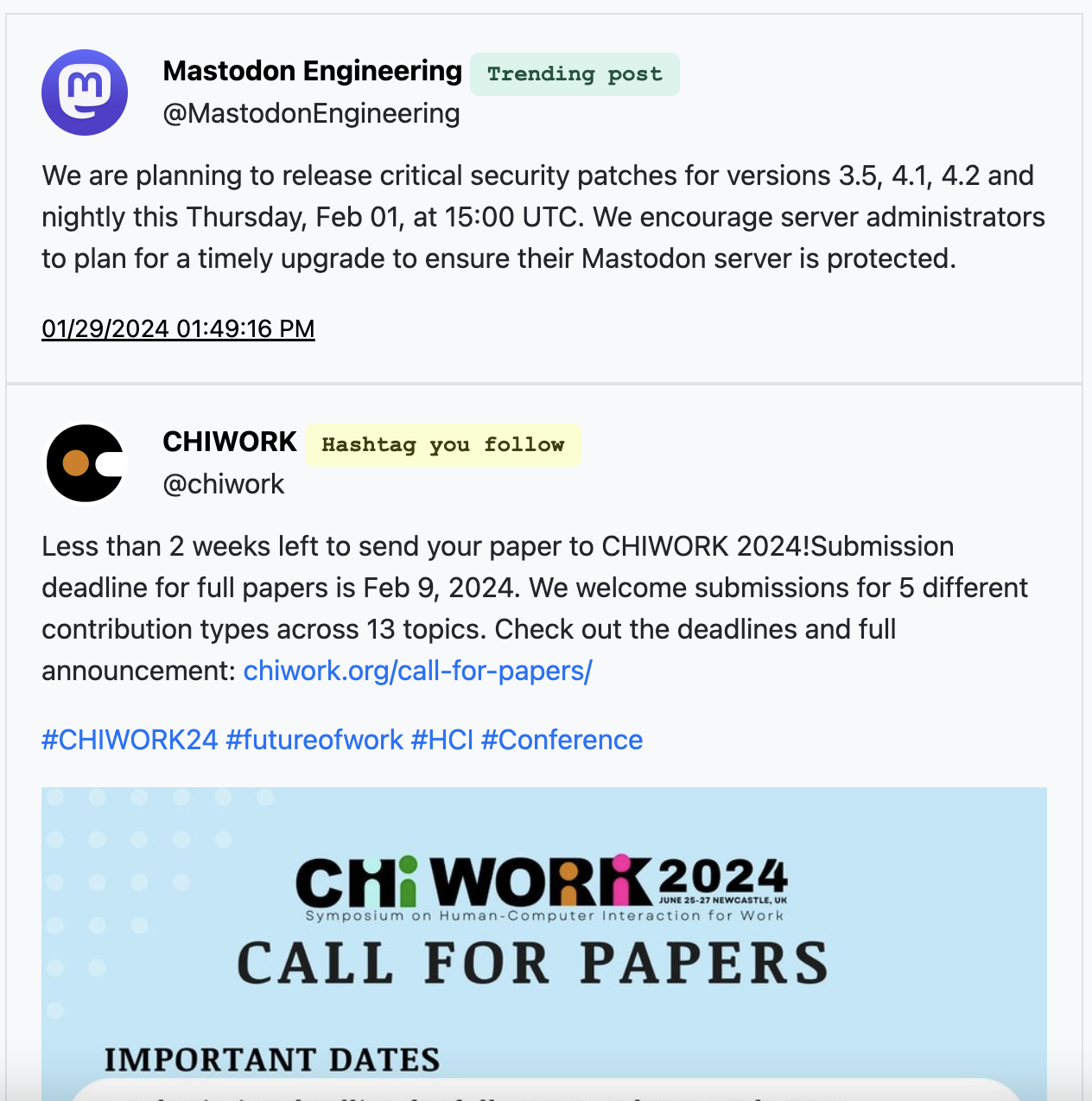}
        \caption{Posts on the unified feed feature badges indicating which feed we pulled it from}
        \label{fig:timeline}
    \end{subfigure}
    \hfill
    \begin{subfigure}[t]{0.32\textwidth}
        \centering
        \includegraphics[width=\linewidth]{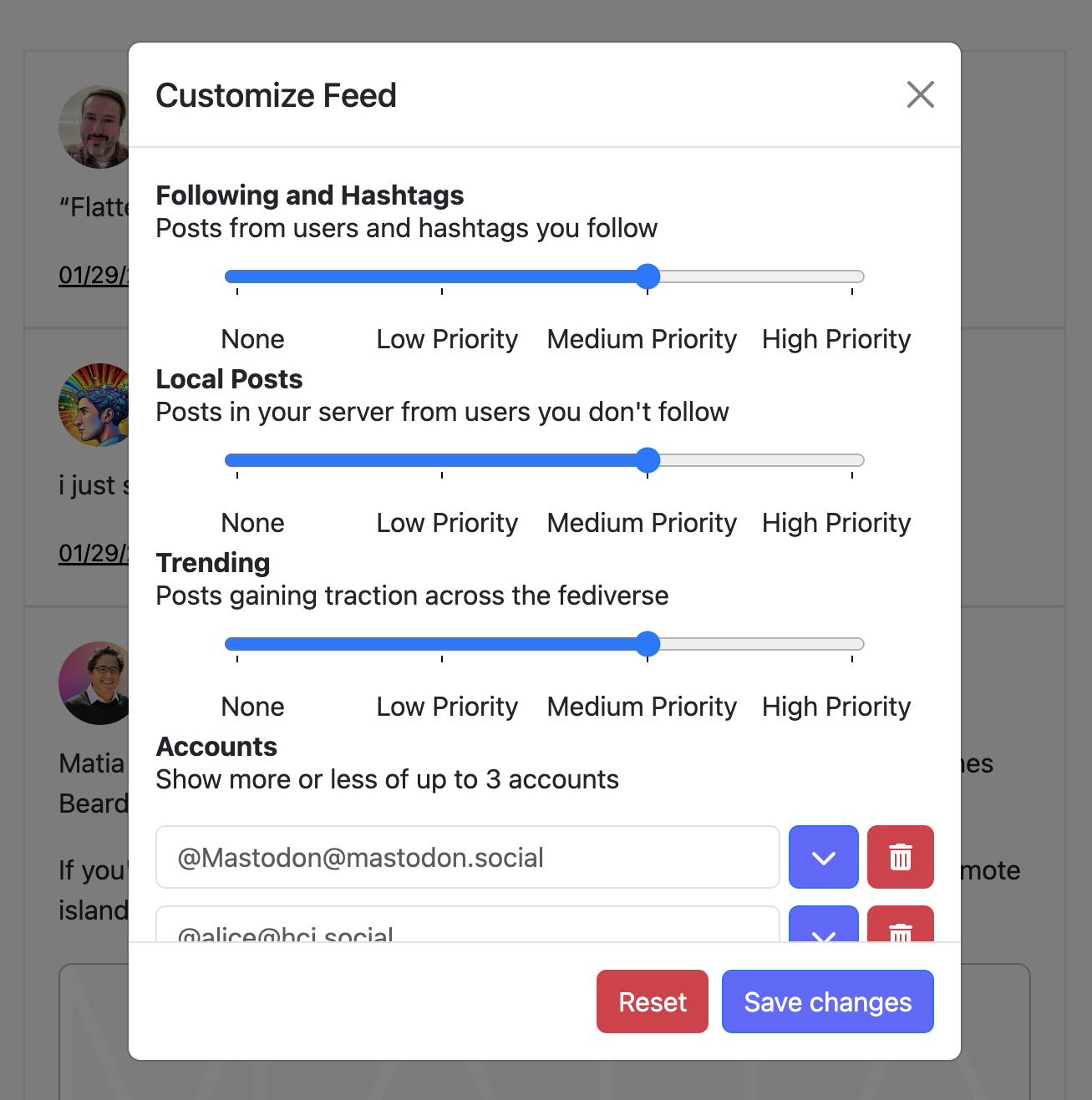}
        \caption{\Braids's default user interface for inputting curation settings}
        \label{fig:interface}
    \end{subfigure}
    \hfill
    \begin{subfigure}[t]{0.32\textwidth}
        \centering
        \includegraphics[width=\linewidth]{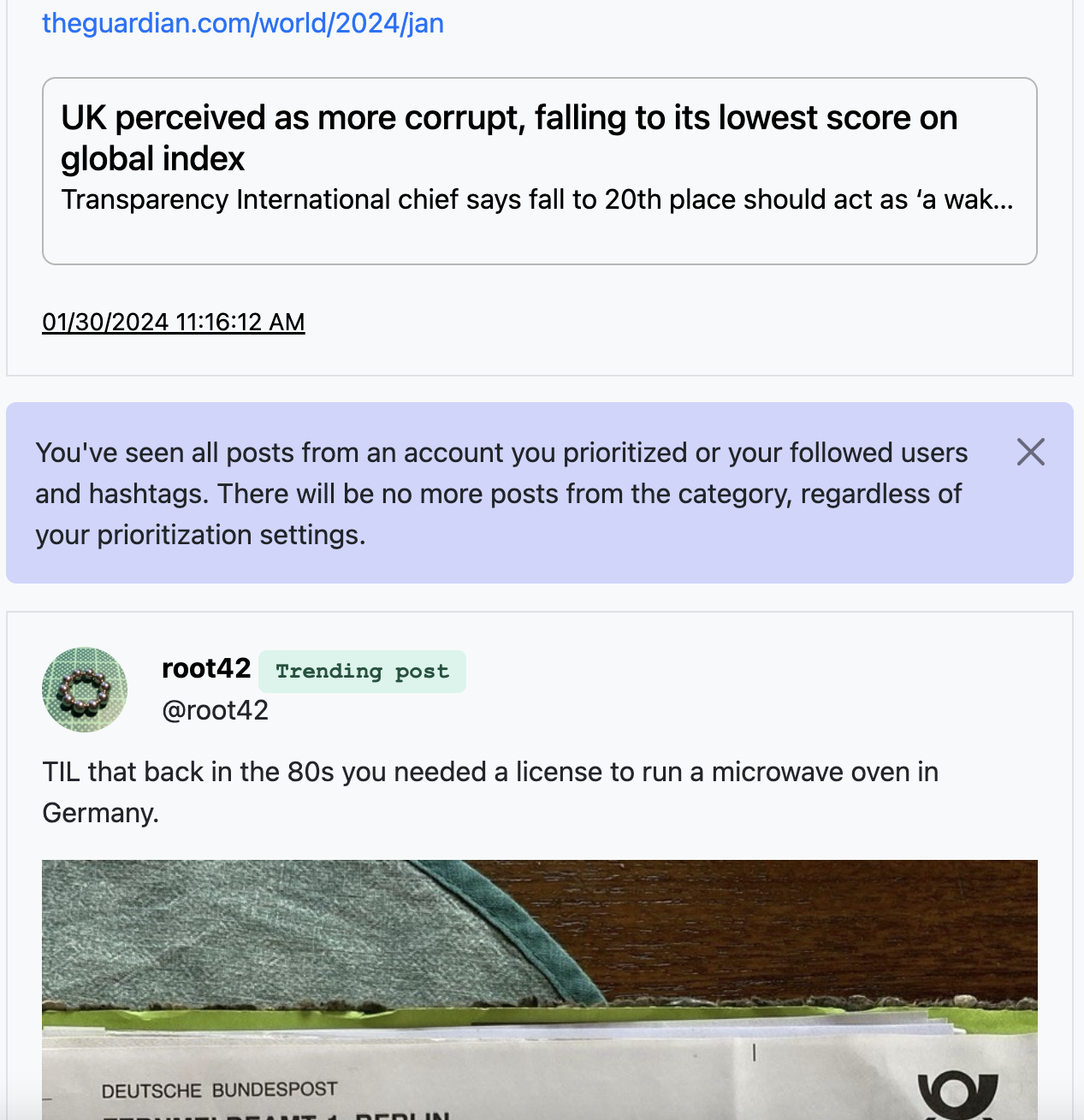}
        \caption{Warning shown running out of posts from one data source}
        \label{fig:ranout}
    \end{subfigure}
    \caption{Design Considerations of \Braids}
    \label{fig:combined}
\end{figure}

\addedcontent{From part one study, we identified two implications in designing feed curation tools in Mastodon: }
\textcolor{black}{
\begin{itemize}
    \item \textbf{Collapsing Posts into a Single Feed to Reduce Information Overload in Local and Federated Feeds:} Participants from part one of the study reported under-utilization of the local and federated feeds due to the overwhelming amount of information. In the meantime, participants also acknowledged the importance of these feeds in providing access to content from a broader network of accounts and topics. Therefore, we designed \Braids with unified feeds where posts are extracted from three main categories: content users directly followed (hashtags and accounts), content from people from the same community, and content from the broader \dquote{fediverse}. 
    \item \textbf{Retain Chronological Order and Indicate Data Source to Increase the Curation Transparency:} Participants from part one study preferred the chronological order and attributed it to its transparency. When designing \Braids, we retained the chronological order within each feed's fragmentation but kept the composition of the unified feed based on the user's settings. In addition, as users value transparency in the algorithmic curation, each post on the generated feed features a badge indicating the data source it was fetched from (See Figure \ref{fig:timeline}). Badges we designed include \dquote{Users you follow}, \dquote{Hashtag you follow}, \dquote{Trending post}, \dquote{Local post} and \dquote{Prioritized account}. The badge only reflects the first feed it was seen in to prevent duplication. 
\end{itemize}
}

\noindent\deletedcontent{\textit{5.1 \quad Collapsing Posts Into A Single Semi-Chronological Feed} The part one study found that participants under-utilized the federated feed due to the overwhelming amount of information it presented. However, participants acknowledged the importance of these feeds in providing access to content from a broader network of users and topics. Therefore, to enhance browsing efficiency, \Braids offers a unified feed that integrates the home, local, and explore feeds, allowing users to customize the composition of each. Each time the page loads or the user clicks `Show more', the system fetches 40 new posts from calls to Mastodon's API. The number of posts fetched from each feed depends on the user's prioritization levels, and we fetch posts proportionate to these levels. The pseudocode for the implementation is found in Appendix \ref{app:code}.}

\deletedcontent{Participants showed a preference for chronological order in the part one study due to the transparency it provides. Therefore, when implementing the system, we tried to retain the chronological order of posts. However, there are more recent posts from the broader local server and remote servers than from the small circle of accounts and hashtags a user follows, which results in the posts from the Home feed typically landing at the bottom. Thus, if a user sets their 'Following and Hashtags' category to `High Priority' and `Trending' category to `Low Priority,' the first posts in the unified feed will all be from the Home feed.}

\deletedcontent{Our approach to combining the feeds is to keep the posts \textit{within} a single API call chronological, but intersperse the feeds together. We do this by treating the posts fetched from one API call as a queue of chronological posts. Until all of the queues are empty, we randomly choose one of the queues (weighted by the prioritization levels), remove its first post, and add it to the end of the unified feed.}

\deletedcontent{A consequence of this chronological ordering, however, is that we will pull older and older posts from the Home feed and specific accounts compared to much more recent posts from the local server and Trending feed. At one point, we may even run out of Home and account posts, in which case we release an alert (Fig. \ref{fig:ranout}) that no further posts from the Home feed or prioritized accounts will be shown, even if the user has assigned it a high prioritization level.}

\subsection{Interface \addedcontent{Design }\deletedcontent{for Collecting User Preferences}}
 \addedcontent{Recent work has validated the efficiency of slider-based design in enhancing transparency in recommendation systems and content moderation tools~\cite{jhaver2023personalizing, he2016interactive, Warren2021IntelBleep}. Inspired by this, we designed a user interface to collect users' preference of how their unified feeds in \Braids{} are composed }(See Figure \ref{fig:interface}). \deletedcontent{\Braids{} provides an interface for collecting user preferences } Through sliders with values ranging from \dquote{None} to \dquote{High Priority}, users indicate how much they want to see content from three \deletedcontent{categories} \addedcontent{data sources}: posts from accounts and hashtags you follow \addedcontent{(directly followed content)}, posts in the user's local server but from accounts they don't follow \addedcontent{(posts from people in the same community)}, and posts gaining traction across the federated remote servers \addedcontent{(posts in the broader \dquote{fediverse})}. \deletedcontent{These categories align with Mastodon's feeds: Home, Local, and Explore.} \addedcontent{Moreover, we displayed the generated feed based on the user's setting but kept the chronological order within each fragment. That is, }if a user set \dquote{High Priority} to home feeds and \dquote{Low Priority} to local feeds, the post fetched from each feed will be displayed chronologically. Still, posts from the home feed will be placed higher than local ones, regardless of posting time. Finally, \addedcontent{as pointed out by prior work, individuals often prioritize content from those they care about~\cite{feng2023teachable}}. We allow people to prioritize specific accounts from any Mastodon instance, regardless of whether they follow them.

\deletedcontent{We use the slider intervals to calculate what proportion of the unified feed a single category takes up. The prioritization levels map onto an ordinal scale: `None' to 0, `Low Priority' to 1, `Medium Priority' to 2, and `High Priority' to 3. For example, setting the priority level of the three feeds to `None' and an account to any non-None priority level creates a feed of a single user's posts. Setting all categories to `High Priority' will create a feed where each category appears in equal proportion.}
% \subsubsection{Increasing Transparency in Feed Curation}

\noindent\deletedcontent{\textit{5.3 \quad Insights from Evaluating Design Features} In the collection of user preferences, we highlighted four key features in prioritizing ranking posts: time, popularity, topics, and account. Each post on the generated feed features a badge indicating which data source it was fetched from (Fig. \ref{fig:timeline}). For each post fetched from the Home feed, we check if the user follows its author. If so, the badge reads `User you follow,' otherwise, `Hashtag you follow.' The other badges are `Trending post,' `Local post,' and `Prioritized account.' It is possible that a post is fetched by multiple API calls. In that case, to prevent duplicate posts in the feed, the badge only reflects the first feed it was seen in. In addition to the badges, we made an animation explaining how the fetching works and made it available to click on the top of the web page.}

% \begin{figure}[hbt!]
%     \centering
%     \includegraphics[width=0.5\linewidth]{Figures/animation.pdf}
%     \caption{A screenshot of the animation explaining how feed in \Braids{} is generated}
%     \label{fig:animation}
% \end{figure}

\subsection{\addedcontent{System Implementation}}
\addedcontent{\Braids{} is a Mastodon web client built with Flask as the backend and React as the frontend. It utilizes the Mastodon API to fetch posts from various data sources: Home Timeline API, Public API (with local parameters), and Trending Status API\footnote{In Mastodon API, status means posts, trending posts feed is a re-ranking of federated feed based on popularity and time of the posts} based on users' configuration, with each request retrieving 40 posts. The user's configuration determines the number of posts fetched from each source. Pseudocode for the implementation is provided in Appendix \ref{app:code}. To use \Braids{}, users must authenticate with their preferred Mastodon account via OAuth 2.0 and grant read-only permissions.}

\section{\addedcontent{Findings: Exploring How to Enable Users to Explicitly Indicate their Preferences in Mastodon Feed Curation} \deletedcontent{RQ3: What features might address people's need for feed curation in Mastodon?}}

\subsection{The Unified Feed Makes Information More Digestible While Adding Discoverability}
First, part two of the study further confirmed the findings from part one of the study that feeds other than home feed are underutilized. When left to configure their feeds, nine out of ten participants (except P18) put the following content (Home feed) as a high priority while setting the others as none, low, or medium, the details of their settings are shown in Table \ref{tab:evaluation_results}. On the one hand, participants viewed the combination of different feeds \pquote{a controlled way to digest local and trending timelines} (P13) because posts from those two feeds are scattered in the content users are more familiar with. On the other hand, participants found the unified feed enabled them to discover new stuff other than what they already followed: \pquote{Definitely like to have a single feed. I mostly ignore other features that are not my main (Home) feed, so, being able to have one feed and mix in the amount of other stuff that I'm interested in seems really attractive to me,} (P15)
\begin{table}[t!]
  \caption{Summary of Feed Customization}
  \label{tab:evaluation_results}
\begin{tabular}{lllll}
\toprule
  ID    &   Following Content  & Local Posts  & Trending Posts  & Accounts \\
\midrule
  P12    & High      & None      & Medium    & None     \\
  P13    & High      & Low       & Low       & None     \\
  P14    & High      & Medium    & None      & None     \\
  P15    & High      & None      & Low       & None     \\
  P16    & High      & Low       & Low       & None     \\
  P17    & High      & Low       & Medium    & None     \\
  P18    & Medium    & None      & Medium    & None     \\
  P19    & High      & Low       & Medium    & None     \\
  P20    & High      & None      & None      & High     \\
  P21   & High      & Low       & None      & None     \\
\bottomrule
\end{tabular}
\end{table}

\subsection{The Adaptable Customization Accommodates Users' Changing Preferences}
In part one of the study, participants identified scenarios where they preferred algorithmic feeds, mainly when they wanted a quick catch-up. While trying out \Braids{}, participants found the customization interface especially helpful for prioritizing what they tried to catch up on. For example, P15 appreciated the design of the priority sliders, stating: \pquote{I can put everything down to None except for the one thing I wanted to make sure I saw.} Similarly, P21 found the \dquote{Prioritized account} feature valuable: \pquote{The fact that I can type in 3 accounts at the bottom to make sure I never miss anything from those accounts kinda gives me another level of comfort of not feeling like I'm gonna miss out on something.} This finding contrasts slightly with our results from the part one study, where participants preferred chronological feeds to fully catch up on missed content during offline periods, rather than relying on algorithmic curation. It is important to note that, in this context, the "algorithms" participants referred to are machine learning-based algorithms commonly used in mainstream social media platforms. We attribute this attitude difference to the transparency and user agency embedded in the design of \Braids. 

However, participants acknowledged that their needs would change over time. Beyond catching up, they expressed interest in adjusting customization settings for other situations. As P14 noted: \pquote{I like the idea of being able to customize what I'm seeing on a you know, daily, weekly basis. If there's some breaking news this week, as far as changing my [preferences]. If I was on an art server, I might be following an artist who's working on a project.} P15 also echoed this: \pquote{I always felt like there were times when I might like to hear about the people that I know personally and have a strong connection to, and everybody else I don't care about. Or maybe I'm in an exploratory mood, and I'd like to see the local and trending stuff and be able to broaden my network that way. So that's interesting, being able to set what I like, and then being able to change it if I want a different mix.} P20 mentioned a desire to toggle back to chronological feeds for more casual browsing: \pquote{Maybe I'm sitting down for an hour of miscellaneous computer time, and that wouldn't be the point where I'd want to use \Braids, because I want the richer data experience.} This finding underscores the importance of flexibility in the tool to accommodate varying user needs while highlighting the value of allowing users to switch off algorithmic curation and regain control over their feed.

\subsection{Enhancing Curation Transparency -- Ensuring Precise Interpretation of Affordances and Continuously Reinforcing Understanding of the Algorithm}

During the part two study, we asked the participants to complete three feed configuration tasks to make sure they understood how to use the system. Most participants found the slider design intuitive and easy to understand. However, we observed that it took extra cognitive effort to understand the differences between sliders’ prioritization levels. For example, P16 said \pquote{It’s intuitive in as much as what [low, medium, high] means. As to how that actually translates, if I really wanted to know, I’d go in there and start plugging away and doing experiments.} In addition, participants developed different interpretations of what \dquote{high, medium, and low} means. In task three, to see twice as many posts from a specific user as trending posts, seven out of ten participants set the categories to high and low priority, respectively, with the other three choosing high and medium priority, respectively. P18 explained his decision to choose the former: \pquote{I think medium will be a more incremental relative increase than high. So the high priority made a very noticeable difference to the low priority. But medium relative to low sometimes it felt like it wasn't really too much of an appreciable difference.}

Moreover, participants identified a couple of \Braids{}'s features as useful indicators that the rules-based algorithm worked, i.e., their input to the prioritization interface changed the feed as expected. All participants found the post badges helpful in seeing where each post was coming from, and as verification, the sliders worked. P14 added, \pquote{I like the feeling that [\Braids{}] responds immediately. Redrawing the timeline when I change the customization rules is really nice. A lot of times you change the rules, and it's like, well, maybe that'll take some effect later. Maybe not. Who knows?} In addition, some participants wanted confirmation that the algorithm worked as expected by seeing the feed's ordering reflect their prioritization settings. In Task 1, to see almost entirely trending posts, P13 set Trending to high priority, Local to none, and Following and Hashtags to low priority and said, \pquote{I started to see a few `Users I follow' all in a row, and I was like, wait a minute. Where'd the trending posts go? As we went further down, most of these say \dquote{Trending post} And I was like, okay, never mind. It's still good.} P13 expected the category with the highest prioritization to appear immediately at the top of the feed, while our random integration of the different timelines does not guarantee this result. The feed should be ordered so that the highest-prioritized posts are seen at the top and consistently throughout to match people's expectations. 
\section{Discussion}
In this section, we reflect on the results of the study and discuss design implications for future curation tools in Mastodon. We distill these into three key takeaways.

% key insights summary
% prior work similarity and difference
% design implication: for future research -- other decentralized realm

\subsection{Seamful Design Enhances People's Acceptance of Algorithmic \addedcontent{Feed }Curation but Faces Challenges in Inaccurate Interpretations}
% explain findings before synthesizing with prior work
Prior work \addedcontent{in mainstream centralized social media platforms} has documented the resistance of algorithmic curation \addedcontent{in social feeds} due to various reasons such as violation of expectations and misunderstanding of the algorithm~\cite{devito2017algorithms, rader2015understanding, eslami2015always, eslami2016first}. However, this resistance was not rooted in dissatisfaction with the curated content. Instead, it stemmed from a sense of \dquote{betrayal} experienced by users, primarily due to their initial unawareness of the algorithm's role or a sense of confusion regarding its operational mechanics~\cite{eslami2015always, eslami2016first}. Recent research has highlighted the efficacy of seamful design approaches in making the algorithms' presence known and enhancing interactions between humans and algorithms ~\cite{eslami2016first, ehsan2024seamful}. Our investigation reveals a consistent trend: users, thanks to seamful design, are cognizant of the algorithmic influence on their content feeds. 

Our findings indicate that while visual cues and real-time feedback improved transparency in algorithmic \addedcontent{feed} curation, participants remained uncertain about precisely interpreting the slider affordances, particularly in understanding the exact impact these settings had on feed generation. As noted by previous research~\cite{jhaver2023personalizing}, providing greater granularity in conveying differences between levels could further enhance users' understanding. We advocate providing more granularity in seamful design and real-time visual feedback to align with users' mental models in interpreting feed curation affordances but leave it to future work to explore the optimal level of granularity. \addedcontent{Moreover, Mastodon is not the only decentralized social media platform that supports multiple feeds. For example, BlueSky allows users to follow various feeds curated by others based on their interests in specific topics and accounts~\cite{bluesky}. The insights gained from our study on seamful design and slider-based interfaces are also applicable to configuring feeds in other decentralized social media platforms. As decentralized networks evolve, designing intuitive, user-controlled curation mechanisms will be critical for enhancing content discoverability while maintaining transparency and user autonomy. Future research should explore how different decentralized platforms implement feed customization and how varying degrees of algorithmic assistance impact user engagement and trust.}

%%%%%%%%%%%%%%%%%%%%%%%%%%%%%%%%%%%%%%%%%%%%%%%%%%%%%%%%%
\subsection{Machine Learning vs. Rule-based Algorithmic Curation}

% rule-based vs. Machine learning

Our study revealed that participants are open to algorithmic curation in addition to their love of chronological feeds. They identified scenarios where a chronological feed cannot be replaced by algorithmic feeds, either machine learning-based or rule-based, and vice versa. In summary, our study results suggest that the feed curation users prefer is closely tied to their browsing intent. Participants appreciated the randomness of chronological feeds for mindless browsing and valued the transparency they offered when they had enough time to catch up on all missed content. However, participants relied on algorithmic assistance when the goal was to catch up on specific topics or individuals quickly. Rule-based algorithms were considered more useful for configuring the characteristics of posts to focus on, while machine learning-based algorithms were better suited for generating quick summaries to help users grasp key information efficiently. \addedcontent{In addition, machine learning-based algorithms are widely recognized to enable personalized content discovery beyond direct subscriptions~\cite{portugal2018use, bhatt2014review}. Recent research on Mastodon administrators' governance challenges highlights the trade-off between transparency in chronological feeds and AI-driven recommendations. The absence of algorithmic content discovery has been identified as a key limitation, posing challenges for user adoption and engagement within Mastodon instances~\cite{zhang2024trouble}}. \addedcontent{In the area of broader recommendation systems, content discovery is also explored with the definition of \dquote{serendipity}(relevance, novelty, and unexpectedness in the recommendation outputs)~\cite{kotkov2016survey, ziarani2021serendipity, reviglio2019serendipity, kotkov2023rethinking}. Previously, both rule-based filters and machine learning methods were proposed to encourage serendipitous discovery while balancing the user's browsing goals in social media, e-commerce, and news~\cite{fletcher2018automated, ge2010beyond, chiu2011social, kotkov2020does}. In this work, we highlighted the opportunity in future research to balance the \dquote{serendipity} and accuracy in UX design in recommendation systems, especially for increasing the engagement and retention of newcomers.}

Even though some participants hesitate to use machine learning algorithms due to their black-box nature, it's important to clarify that it does not necessarily indicate a general opposition to them. Considering that currently, most Mastodon servers \addedcontent{and servers in decentralized social media} operate as non-profit and ad-free entities, it remains an open question whether user attitudes would shift over time, especially in light of recent advancements in the transparency and explainability of machine learning algorithm curation on social media ~\cite{gil2022effects, ridley2023using, ngo2022exploring}. This evolving landscape presents several opportunities for future research. First, there is a need to explore in-depth the user perceptions of machine learning-based algorithmic curation within the Mastodon network. Second, we see potential in co-designing the combination of rule-based and machine learning-based algorithms, actively involving Mastodon users in the development process. Such research could provide valuable insights into user preferences and contribute to creating more user-centric algorithmic solutions in decentralized social media.
%%%%%%%%%%%%%%%%%%%%%%%%%%%%%%%%%%%%%%%%%%%%%%%%%%%%%%

\subsection{Development Choice in the Fediverse: New Apps or Add-ons?}
Another observation from our study is that many participants had experience using third-party apps, with some even primarily relying on them. The popularity of third-party apps can be partially attributed to the ActivityPub protocol, which allows users to access interoperable data across the federation network without installing separate apps for each platform. Participants in our studies normally use more than one app as no singular solution can fully cater to all their needs. While the decentralized nature of the Fediverse allows for various clients, each addressing specific user requirements, we observed a reluctance among users to download multiple apps for different features. The tension between having multiple apps and better user experiences resonates with the advocacy made by Masnick about \textit{build protocols, not platforms}. \addedcontent{The dissatisfaction of policies on a single platform pushes people to start more and more platforms. However, these new platforms cannot satisfy users' expectations all at once and lack interoperability~\cite{masnick2019protocol}.} \deletedcontent{where the openness of the protocol will encourage tech innovations} With users gaining more freedom to choose which interface or app they use to interact with protocol-based social media, the UX challenge remains similar to that of managing multiple platforms: users still need to navigate multiple apps for different features, much like managing different accounts across different platforms. Therefore, we raised the question here whether app developers should open access to make add-on features possible for other developers. We suggest future research to investigate the trade-offs between creating new apps versus developing add-ons from the perspectives of both users and developers.

\section{Limitations}
Our work has several limitations. First, our findings are limited by the small sample size \deletedcontent{, which may not represent all Mastodon users. Moreover,}. Mastodon users are generally more concerned about algorithmic transparency, anti-corporation, and privacy than traditional social media users \deletedcontent{, which may result in biased insights}. Our use of Google Forms for recruitment and Zoom for interviews may \deletedcontent{have} deterred \addedcontent{some} \deletedcontent{highly anti-corporate} Mastodon users from participating in the study. Thus, our insights may not represent all Mastodon users, who are not representative of all social media users.

\addedcontent{Furthermore, Mastodon is a microblogging platform within the broader decentralized social media ecosystem, alongside platforms like PeerTube for video-sharing and Pixelfed for photo-sharing. These platform-specific focuses may shape different user expectations, particularly in how users interact with strong and weak ties and discover new content. As a result, our findings on feed curation in Mastodon may have limited generalizability across other decentralized platforms with different interaction dynamics. Additionally, Mastodon operates on the ActivityPub protocol~\cite{activitypub}, which differs in decentralized architecture and information flow from other protocols such as ATProtocol~\cite{bluesky} and Farcaster~\cite{farcasterdocs}. Consequently, the insights gathered from our user study may not be directly transferable to platforms using different underlying protocols without significant adaptation.}

\section{Conclusion}
The decentralized social media Mastodon has attracted considerable attention after Twitter announced a series of policy changes. In this paper, we conducted a two-part interview study with Mastodon users to understand how they perceive, interact with, and manage Mastodon feeds and investigate how we can map the user preference in developing tools in Mastodon feed curation. The study filled the gap in understanding the dynamics of feed curation in decentralized social media. Our findings highlight several areas for future HCI research. First, we noticed that seamful design can enhance people's acceptance of algorithmic curation but also faces challenges in conveying affordance more accurately. Second, we discussed the tension between machine learning-based and rule-based algorithms in feed curation, highlighting that both methods have irreplaceable use cases. Finally, we advocate for future HCI researchers to investigate the pros and cons of developing new applications versus integrating add-ons within the Fediverse ecosystem. 

\section{Disclosure of the usage of LLM}
We used ChatGPT (GPT4o model\cite{achiam2023gpt}) to facilitate the writing of this manuscript. The usage includes:
\begin{itemize}
    \item Turn Excel format tables into LaTeX format tables
    \item Correct grammar mistakes and spelling
    \item Polish the existing writing by prompts like ``Find me a synonym of X,'' ``What is the noun/adjective form of X'' and ``Shorten this sentence without changing its content.''
\end{itemize}

%%
%% The acknowledgments section is defined using the "acks" environment
%% (and NOT an unnumbered section). This ensures the proper
%% identification of the section in the article metadata, and the
%% consistent spelling of the heading.
\begin{acks}
We would like to thank Laiba Ali for her support in conducting interviews and generating transcripts. We would also like to thank all our interviewees for their insights.
\end{acks}

%%
%% The next two lines define the bibliography style to be used, and
%% the bibliography file.
\bibliographystyle{ACM-Reference-Format}
\bibliography{0_ref}

%%
%% If your work has an appendix, this is the place to put it.
\appendix
\section{Basic Information of Study Participants}\label{tab:merged_participants}
To exclude personal information, we use general descriptions of their home server instead of explicitly listing the exact ones. \\

\label{appendix:participants}
\begin{table}[H]
  \caption{Basic Information of Participants in the Exploratory and User Study}
  
\begin{tabular}{llllll}
\toprule
    ID   & Year Joined & Gender & Home Server               & Country & Mastodon Client \\
    \midrule
    P1   & 2020        & Female & Academic                  & USA     & Native app, Elk \\
    P2   & 2023        & Female & Art creators             & USA     & TweeseCase, Mona, Tuskey \\
    P3   & 2021        & Male   & Anarchist and Communist  & USA     & Native app, Tusky \\
    P4   & 2022        & Female & General-purpose          & USA     & Native app \\
    P5   & 2022        & Female & General-purpose          & USA     & Native app \\
    P6   & 2022        & Male   & General-purpose          & USA     & Ivory \\
    P7   & 2022        & Female & Regional                 & USA     & Native app \\
    P8   & 2023        & Male   & General-purpose          & USA     & Fedilab \\
    P9   & 2022        & Female & People of color          & USA     & Native app, justmytoots.com \\
    P10  & 2018        & Male   & Regional                 & USA     & Elk, Fedilab \\
    P11  & 2016        & Female & General-purpose          & USA     & Native app \\
    P12  & 2022        & Female & Not Specified            & USA     & Native app, Tusky \\
    P13  & 2022        & Male   & Not Specified            & USA     & Elk, Ivory, Ice Cubes \\
    P14  & 2022        & Male   & Not Specified            & USA     & Native app, Mastonaut \\
    P15  & 2022        & Male   & Not Specified            & USA     & Native app, Fedilabs \\
    P16  & 2018        & Male   & Not Specified            & Canada  & Elk, Megaladon, Tuba \\
    P17  & 2022        & Male   & Not Specified            & USA     & Native app, Ivory, Megalodon \\
    P18  & 2023        & Male   & Not Specified            & USA     & Native app, Moshido \\
    P19  & 2017        & Female & Not Specified            & USA     & Native app, Tusky \\
    P20  & 2018        & Male   & Not Specified            & Germany & Native app, Ice Cubes \\
    P21  & 2017        & Male   & Not Specified            & USA     & Native app, Ivory \\
\bottomrule
\end{tabular}
\end{table}

\section{The Rules-based Curation Algorithm of \Braids}\label{app:code}

Below is the pseudocode for the two functions used to generate the unified feed given the user's input to the prioritization interface. We call the Mastodon API in \texttt{getFeed()} to retrieve the posts. We combine the posts from the different timelines in \texttt{combinePosts()}.

\begin{verbatim}
# This function takes the posts retrieved from the different feeds and merges them together based 
on the prioritization level assigned to each feed
def combinePosts(posts):
  
  seen = dictionary of IDs of posts we have already
  seen (to prevent duplicates)
  
  scoreSum = total sum of the prioritization levels
  
  filters = get list of content filters

  if we have posts from the Home feed:
    map each post from Home to whether or not its  author is a followed user

  while posts is non-empty:
    randomly select one of the non-empty categories of posts
    
    remove the category's most recent post from the list of posts
    
    remove category if it is now empty

    if the post has not been seen already and does not contain a filtered-out phrase:
        reformat and add a field for post category to the post's JSON
        
        add this post's id to be seen
        
        break
      
    return the list of seen posts

# This function retrieves posts from Mastodon's feeds via the API in proportion to the user's 
prioritization specifications
# scores is a dictionary with category names as keys, and the integer mapping of the prioritization levels
as values (none is 0, low is 1, medium is 2, high is 3) 
# first_page is a boolean that is True when we have just reloaded the page, and False if we have clicked
'Show more' to see new posts
# Returns a tuple: (list of post JSONs, ranOut: boolean for whether or not to display the warning that we've 
run out of posts from either Home feed or a prioritized account)

def getFeed(scores, first_page):

    last_index = dictionary mapping categories to the
    index of the last post retrieved
  
    ranOut = False

    reset last_index to initial values if first_page is True

    sum = sum of the prioritization scores
    if sum is 0:
        return (empty list of posts, False)

    # Function retrieving posts from finite feeds such as Home feed or prioritized accounts
    
    def getFinitePosts():
    
        if the category received prioritization level None
        or we've run out of posts:
            return empty list of posts
        
        calculate proportional number of posts to retrieve based on prioritization level: 
        (score * 40) // sum
        
        if the category is Home:
            retrieve appropriate number of posts from the Home feed via Mastodon API
        else:
            retrieve appropriate number of posts from the prioritized account
        
        if 0 posts were returned:
          ranOut = True
          return empty list of posts
          
        update last_index[category]
        return posts

    # Function retrieving posts from infinite feeds like Local and Trending
    
    def getInfinitePosts():
    
        if the category received prioritization level None:
            return empty list of posts

        calculate proportional number of posts to retrieve based on prioritization level: 
        (score * 40) // sum
            
        if the category is Local:
            retrieve appropriate number of posts from the Local feed via Mastodon API
        else:
            retrieve appropriate number of posts from the Trending feed via Mastodon API
        
        update last_index[category]
        return posts
  
    # Retrieve posts with Mastodon API
    posts = {}
    for each slider category:
        if category is Local or Trending:
            category_posts = getInfinitePosts()
        else:
            category_posts = getFinitePosts()
        if len(category_posts) > 0:
            posts[category] = category_posts

  # Combine the posts from the three timelines all_posts = combinePosts(posts)
  return all_posts, ranOut

\end{verbatim}

\section{Part 1 Codebook}\label{tab:part_one_codebook}

\begin{table}[htbp]
\centering
\caption{Part 1 Codebook}
% \label{tab:part_one_codebook}
\begin{adjustbox}{max width=\textwidth}
\begin{tabular}{lll}
\toprule
Code & Subcode & Definition \\
\midrule
Usage Of Third Party Software & Usage Of Third Party Software & What third-party software participants use. \\
Why Third Party Software & Dislike Official App & Why they use third party software instead of official app. \\
 & Like Third Party App &  \\
Device Use & Device Use & On what device participant use to browse content. \\
Homeserver Selection & Choice Of Homeserver & How participants choose their home servers. \\
 & Server Migration & Participants migrating from another server to their current home server. \\
Reason Joining Mastodon & Reason Joining Mastodon & Why participants decided to join Mastodon. \\
Joining Time & Joining Time & When the participant joined / created an account on Mastodon. \\
Active Time & Active Time & Some people join before 2022 but are not active until 2022 Twitter migration event. \\
Reason Not Active & Reason Not Active & Why participants were not active immediately after joining. \\
Reason Become Active & Reason Become Active & Why they became active after joining. \\
Usage Similarity Between Twitter And Mastodon & Usage Similarity Between Twitter And Mastodon & Similar behavior on Twitter vs. Mastodon. \\
Usage Difference Between Twitter And Mastodon & Usage Difference Between Twitter And Mastodon & Different behavior on Twitter vs. Mastodon. \\
Community Motivation Behind Mastodon Use & Community Motivation Behind Mastodon Use & Participant's motivation to join their homeserver. \\
Frequency Checking Home & Frequency Checking Home & How often they check home timeline. \\
Frequency Checking Local & Frequency Checking Local & How often they check local timeline. \\
Frequency Checking Federated & Frequency Checking Federated & How often they check federated timeline. \\
Frequency Checking Explore & Frequency Checking Explore & How often they check explore timeline. \\
Reasons Of Frequency Difference & Reasons Of Frequency Difference & Why do they check one timeline more frequent than others? \\
Curation Of Home & Discover New Follows & What methods do they use to curate their home timeline. \\
 & Follow Hashtags &  \\
Timline & Timline & Whether the timeline is related/aligned with user's motivation/goals. \\
Like Of Chronological Order & Like Of Chronological Order & Participant likes chronological order. \\
Dislike Of Chronological Order & Dislike Of Chronological Order & Participant dislikes chronological order. \\
Like Of Algorithmic Mediation & Like Of Algorithmic Mediation & Participation likes the idea of algorithmic feed. \\
Dislike Of Algorithmic Mediation & Dislike Of Algorithmic Mediation & Participation dislikes the idea of algorithmic mediation. \\
Like Of Mastodon User Base & Like Of Mastodon User Base & Participant likes the userbase of Mastodon, such as nice people, friendly community, better quality posts. \\
Dislike Of Mastodon User Base & Dislike Of Mastodon User Base & Participant disliked the userbase of Mastodon, such as political extremism, low quality content. \\
Confusion About Mastodon Terms & Confusion About Mastodon Terms & Participant does not understand Mastodon terminology. \\
Confusion About Ui & Confusion About Ui & Participant gets confused over Mastodon UI, such as multiple timelines and accounts management. \\
Privacy Concern & Privacy Concern & Participant is concerned about privacy of data. \\
Wishlist Of Home & Wishlist Of Home & Features the participant would like on home timeline. \\
Wishlist Of Local & Wishlist Of Local & Features the participant would like on local timeline. \\
Wishlist Of Federated & Wishlist Of Federated & Features the participant would like on federated timeline. \\
Wishlist Of Explore & Wishlist Of Explore & Features the participant would like on explore timeline. \\
General Wishlist & General Wishlist &  \\
Remove Duplicates In Timelines & Remove Duplicates In Timelines & There are duplicate posts in home/local/federated timelines. \\
Like Of Home & Like Of Home & Participant likes home timeline and why. \\
Like Of Local & Like Of Local & Participant likes local timeline and why. \\
Like Of Federated & Like Of Federated & Participant likes federated timeline and why. \\
Like Of Explore & Like Of Explore & Participant likes explore timeline and why. \\
Dislike Of Home & Dislike Of Home & Participant dislikes home timeline and why. \\
Dislike Of Local & Dislike Of Local & Participant dislikes local timeline and why. \\
Dislike Of Federated & Dislike Of Federated & Participant dislikes federated timeline and why. \\
Dislike Of Explore & Dislike Of Explore & Participant dislikes explore timeline and why. \\
Different Interaction On Different Timelines & Different Interaction On Different Timelines & How to participant interact (boost, like, retoot) differently across timelines. \\
Important Post Attribute & Important Post Attribute & Participant expresses what qualities of a post are important to them. \\
Like Of Moderation & Like Of Moderation & Participants likes the moderation of their server. \\
Concerns Of Moderation & Concerns Of Moderation & Participants dislikes the moderation of their server. \\
Mastodon Atmosphere & Mastodon Atmosphere & Participants discusses Mastodon atmosphere and their feelings about it. \\
Usage Of Multiple Accounts & Usage Of Multiple Accounts & Participant discusses their usage of multiple accounts on Mastodon. \\
\bottomrule
\end{tabular}
\end{adjustbox}
\end{table}

\section{Part 2 Codebook}\label{tab:part_two_codebook}

\begin{table}[htbp]
\centering
\caption{Part 2 Codebook}
\begin{adjustbox}{max width=\textwidth}
\begin{tabular}{ll}
\toprule
Code Group & Code \\
\midrule
Mastodon Usage & Primarily prefers the Home feed because they've curated it carefully (personalized content, trusted sources). \\
 & Uses local feed or hashtags when they've run out of content on the Home feed. \\
 & Uses lists to catch up with content on a specific topic or group. \\
 & Doesn't generally like to see trending posts and prefers content they've explicitly chosen. \\
 & Believes Mastodon should have more tools for curating timelines. \\
 & Likes chronological ordering because they don't trust algorithmic feeds. \\
 & Prefers having multiple columns or less white space to see more content at once. \\
 & Wants to reduce the amount of random or suggested content. \\
 & Mastodon needs to work on discoverability. \\
Usefulness of Braids.Social & Seeing content from firehose in a controlled way. \\
 & Discoverability and seeing new content. \\
 & Makes sure they don't miss any content for efficient catching up. \\
 & Adapting to preferences over time or in different contexts. \\
 & A convenient, on-the-go customization tool. \\
 & Maintains a chronological timeline where they still have control. \\
 & Primarily wants to see their Home timeline with subtle adjustments from Braids. \\
 & Wants the flexibility to see their Home feed (via Braids) alongside other content streams. \\
 & Finds prioritizing accounts useful for seeing their favorite content first. \\
 & Seeing local posts is not useful for large instances. \\
 & Braids is not useful for search. \\
 & Prefers a simple, readable interface. \\
 & Miscellaneous. \\
Algorithmic Understanding & Confusing to understand intuitively without explanation. \\
 & Confusing to understand the relationship between the sliders and timeline adjustments. \\
 & Informational section was helpful for understanding algorithmic adjustments. \\
 & Needs greater reinforcement of the informational tooltips. \\
 & Badges were helpful for seeing where each post came from. \\
 & Would want to see multiple badges if it applies to more than one category. \\
 & Would like immediate confirmation of the rules-based outcomes. \\
 & Would want a little more clarity to ensure they understand how sliders affect the timeline. \\
 & Initially believed "Prioritized Accounts" referred to accounts they interact with most often. \\
 & Did not find it obvious how to access the slider interface initially. \\
Slider Interface & Wants to be able to prioritize specific instances. \\
 & Wants to be able to prioritize users and hashtags separately. \\
 & Might enjoy more granularity in the different slider options. \\
 & Sliders provided more control and were easy to adjust quickly. \\
 & More time would be needed to understand the different slider options fully. \\
 & Enjoys having proportion sliders rather than a binary on/off. \\
 & Expressed intuitive understanding of the different sliders after brief use. \\
 & Liked being able to mute or set a category to None. \\
\bottomrule
\end{tabular}
\end{adjustbox}
\end{table}

\end{document}